\documentclass{sigchi-ext}
\usepackage[T1]{fontenc}
\usepackage{textcomp}
\usepackage[scaled=.92]{helvet} % for proper fonts
\usepackage{graphicx} % for EPS use the graphics package instead
\usepackage{balance}  % for useful for balancing the last columns
\usepackage{booktabs} % for pretty table rules
\usepackage{ccicons}  % for Creative Commons citation icons
\usepackage{ragged2e} % for tighter hyphenation
\usepackage[font={small}]{subfig} % subfig, 4 figures in a row

\usepackage{xspace}     % fix space in macros
\usepackage{balance}     % to better equalize the last page
\usepackage{bold-extra}     % bold + {small capital, italic}
\usepackage{siunitx}          % \num for decimal grouping
\usepackage[vlined,linesnumbered,ruled,noend]{algorithm2e}     % algorithms
\usepackage[square,numbers]{natbib}     % better references
\usepackage{microtype}    % compress text
\usepackage{units}     % nicer slanted fractions
\usepackage{mathtools}     % amsmath++

\newcommand{\spara}[1]{\smallskip\noindent\textbf{#1}}

\graphicspath{{img/}}

\title{Exploring Controversy in Twitter}

\numberofauthors{4}
\author{%
  \alignauthor{%
    \textbf{Kiran\\Garimella}\\
    \affaddr{Aalto University} \\
    \affaddr{Helsinki, Finland} \\
    \affaddr{kiran.garimella@aalto.fi} }\alignauthor{%
    \textbf{Michael\\Mathioudakis}\\
    \affaddr{HIIT}\\
    \affaddr{Helsinki, Finland}\\
    \email{michael.mathioudakis@hiit.fi} } \vfil \alignauthor{%
    \textbf{Gianmarco\\De Francisci Morales}\\
    \affaddr{Aalto University}\\
    \affaddr{Helsinki, Finland}\\
    \email{gdfm@acm.org} }\alignauthor{%
    \textbf{Aristides\\Gionis}\\
    \affaddr{Aalto University}\\
    \affaddr{Helsinki, Finland}\\
    \email{aristides.gionis@aalto.fi} } \vfil 
}

\def\plaintitle{Exploring Controversy in Twitter} \def\plainauthor{Kiran Garimella, Gianmarco De Francisci Morales, Michael Mathioudakis, Aristides Gionis}
\def\plainkeywords{Social media; Controversy}

\hypersetup{%
  pdftitle={\plaintitle}, pdfauthor={\plainauthor},
  pdfkeywords={\plainkeywords}, }

\begin{document}

\maketitle

% Uncomment to disable hyphenation (not recommended)
% https://twitter.com/anjirokhan/status/546046683331973120
\RaggedRight{} 
\enlargethispage{1.2\baselineskip}

% Do not change the page size or page settings.
\begin{abstract}
%We demonstrate a system to explore controversy on Twitter.
% Our work is motivated by the observation that, 
Among the topics discussed on social media, some spark more heated debate than others.
For example, experience suggests that major political events, such as a vote for healthcare law in the US, would spark more debate between opposing sides than other events, such as a concert of a popular music band.
Exploring the topics of discussion on Twitter and understanding which ones are controversial is extremely useful for a variety of purposes, such as for journalists to understand what issues divide the public, or for social scientists to understand how controversy is manifested in social interactions.
% Nevertheless, there exist few automated systems to perform this task.

% We present the first system to automate the exploration of controversy on Twitter.
The system we present processes the daily trending topics discussed on the platform, and assigns to each topic a \emph{controversy score}, 
which is computed based on the interactions among Twitter users, 
and a visualization of these interactions, 
which provides an intuitive visual cue regarding the controversy of the topic.
The system also allows users to explore the messages (tweets) associated with each topic, and sort and explore the topics by different criteria (e.g., by controversy score, time, or related keywords). % @Kiran: we do not do related keywords now, could we do that?
\end{abstract}


% \category{See}{\url{http://acm.org/about/class/1998/}}{for   full list of ACM classifiers. This section is required.}

%\input{introduction.tex}
% !TEX root = controversy.tex
\section{Introduction}
\label{sec:intro}

%\enlargethispage{\baselineskip}

% social media, high level motivation
Social media have emerged as the fora of choice for users on the Web to express their opinion about issues they deem important.
These fora provide a way to interact with other users who wish to discuss the same issues.
Due to their widespread adoption, and the fact that much of the activity they host is publicly available, they offer a unique opportunity to study social phenomena such as peer influence, framing, bias, and controversy.
Our work, in particular, is motivated by interest in observing \textbf{controversies} at societal level, monitoring their evolution, and possibly understanding which issues become controversial and why.

% what the system does
The system that we demonstrate focuses on the exploration of controversy on Twitter, currently the most popular micro-blogging platform.\footnote{With 320 million monthly active users as of 30 September 2015 according to \url{https://about.twitter.com/company}.}
The back-end of the system processes the messages generated on the platform on a daily basis in order to ($i$) identify different topics of discussion, ($ii$) assign a controversy score to each topic, and ($iii$) produce visual renderings of the activity surrounding each topic in a way that clarifies whether the topic is controversial.
The front-end provides a web interface\footnote{\url{http://users.ics.aalto.fi/kiran/controversy}} that allows to explore the identified topics according to various views (e.g., ordered by time or magnitude of controversy) and obtain more information about each topic (e.g., by providing a keyword summary of the topic, representative tweets, or a visualization of the activity).

% what is a topic
The system is designed to identify controversy on topics in any domain, i.e., without any prior domain-specific knowledge about the topic in question.
Specifically, topics are defined by \emph{hashtags}, special keywords conventionally employed by Twitter users to signal that their messages belong to a particular topic.
As an example,``\#beefban'' is a hashtag that was employed to convey that a post referred to a decision of the Indian government, in March 2015, about the consumption of beef meat in India. 
The system leverages this convention and treats each hashtag as a different topic.

% how we measure the controversy of each topic
Given a hashtag, we represent the activity on the corresponding topic by a \emph{retweet graph}.
In this graph, vertices represent Twitter users who have used the hashtag at least once on a given day, and edges represent \emph{retweets} between users.
To quantify the controversy of each topic, we rely on the hypothesis that the structure of the retweet graph reveals how controversial the topic is.
This hypothesis is based on the fact that a controversial topic entails 
different sides with opposing points of view, as well as on previous evidence that individuals on the same side tend to endorse and amplify each other's arguments~\cite{adamic2005political, akoglu2014quantifying, conover2011political}.
We studied this hypothesis in previous work~\cite{garimella2016quantifying}, and found strong evidence that the retweet graph of a controversial topic presents a \emph{clustered structure} that reveals the opposing sides of the debate.
Moreover, in the same work, we developed a random-walk-based measure that quantifies accurately how controversial a topic is by taking into account the structure of its retweet graph.
In light of these findings, for each topic identified, the system computes a controversy score, and produces a rendering of the retweet graph that highlights its clustering structure.

%In what follows, we provide further details on the algorithms used to measure the controversy score of a topic (Section~\ref{sec:pipeline}) and visualize its retweet graph (Section~\ref{sec:visualization}), as well as the demonstration that we propose (Section~\ref{sec:demo}).

%\input{related_work.tex} % TODO we can remove or merge with intro
% !TEX root = controversy.tex
\section{Related Work}
Previous studies aim at identifying controversial issues, mostly around political debates~\cite{adamic2005political,conover2011political,mejova2014controversy,morales2015measuring}
but also other topics~\cite{guerra2013measure}.
While most recent papers focus on  Twitter~\cite{conover2011political,guerra2013measure,mejova2014controversy,morales2015measuring},  controversy in other social-media platforms, such as blogs~\cite{adamic2005political} and
opinion fora~\cite{akoglu2014quantifying}, has also been analyzed. 
The main limitation of previous work is that the majority of studies have focused on  known, long-lasting debates, such as elections~\cite{adamic2005political, conover2011political}.
Our system is the first one to attempt controversy detection in the wild, on any topic, and without human data curation~\cite{garimella2016quantifying}.

%\input{pipeline.tex}
% !TEX root = controversy.tex
\section{Quantifying Controversy}
\label{sec:pipeline}

\begin{marginfigure}[0pc]%
\centering%
\includegraphics[width=0.7\marginparwidth]{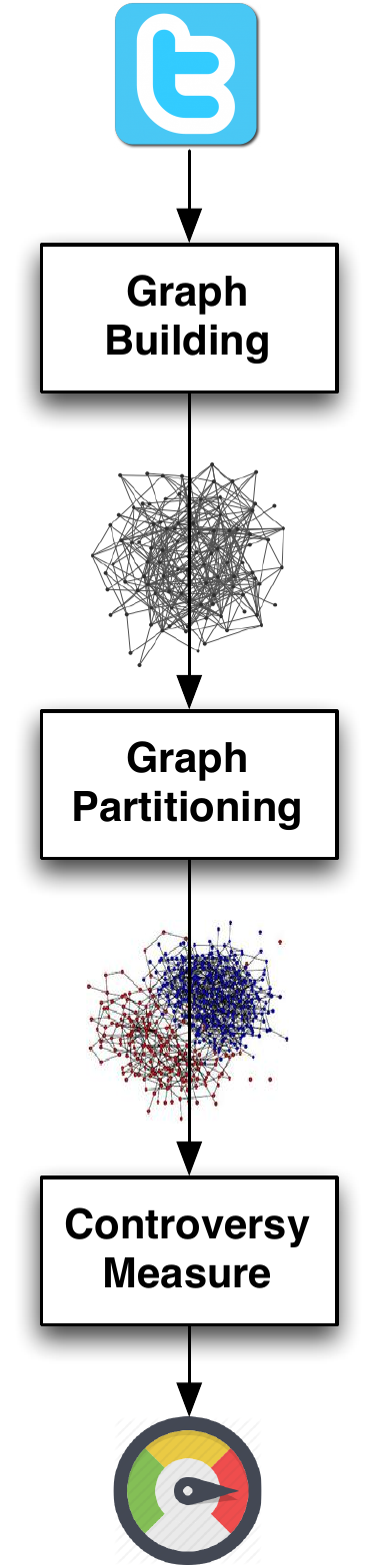}%
\caption{Pipeline for computing controversy scores.}%
\label{fig:block}%
\end{marginfigure}

Our approach to measuring controversy follows a pipeline with three stages, namely \emph{graph building}, \emph{graph 
partitioning}, and \emph{measuring controversy}, as depicted in 
Figure~\ref{fig:block}. The input to the pipeline is a single hashtag, which defines a topic of discussion.
The final output of the pipeline is a value between zero and one that measures how controversial a topic is, with higher values corresponding to higher degree of controversy.
We provide a high-level description of each stage here, for further details refer to the original work~\cite{garimella2016quantifying}.

\subsection{Building the Graph}
\label{sec:pipeline_graph}
The purpose of this stage is to build the \emph{retweet graph} associated with a single \emph{topic} of discussion. 
For a given day, each tweet that contains the hashtag that defines the topic is associated with one user who generated it, and we build a graph where each user who contributed to the topic is assigned to one vertex.
In this graph, an edge between two vertices signifies that there was one retweet between the corresponding users.
We take a retweet as a signal of \emph{endorsement} of opinion between the users.

\subsection{Partitioning the Graph}
\label{sec:pipeline_partition}
In the second stage, the resulting retweet graph is fed into METIS~\cite{karypis1995metis} a \emph{graph partitioning} algorithm to extract \emph{two} partitions.
Intuitively, the two partitions correspond to two disjoint sets of users who possibly belong to different sides in the discussion.
In other words, the output of this stage answers the following question:
``assuming that users are split into two sides according to their point of view on the topic, which are these two sides?''.
If indeed there are two sides which do not agree with each other --a controversy-- then the two partitions should be only loosely connected to each other, given the semantic of edges.

% controversy
\subsection{Measuring Controversy}
The third and last stage takes as input the \emph{retweet graph} built by the first stage and partitioned by the second stage, and computes the value of a random-walk-based \emph{controversy measure}~\cite{garimella2016quantifying} that characterizes how controversial the topic is.
Intuitively, the controversy measure captures how separated the two partitions are.

\section{Visualizing controversy}
\label{sec:visualization}
\begin{marginfigure}[-50pc]
\begin{center}
\includegraphics[width=0.95\marginparwidth]{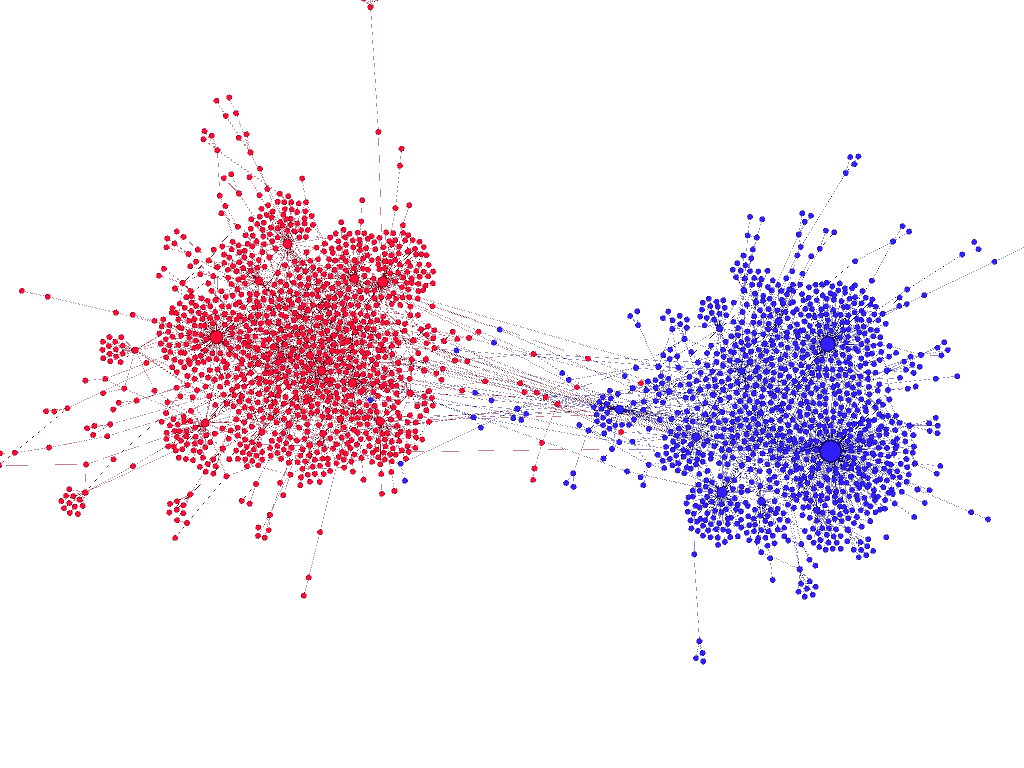}
\caption{Force directed layout visualization of a controversial topic (\#russia\_march).}
\label{fig:russia_march}
\end{center}
\vspace{\baselineskip}
\end{marginfigure}

\begin{marginfigure}[-20pc]%
\begin{center}
\includegraphics[width=0.95\marginparwidth]{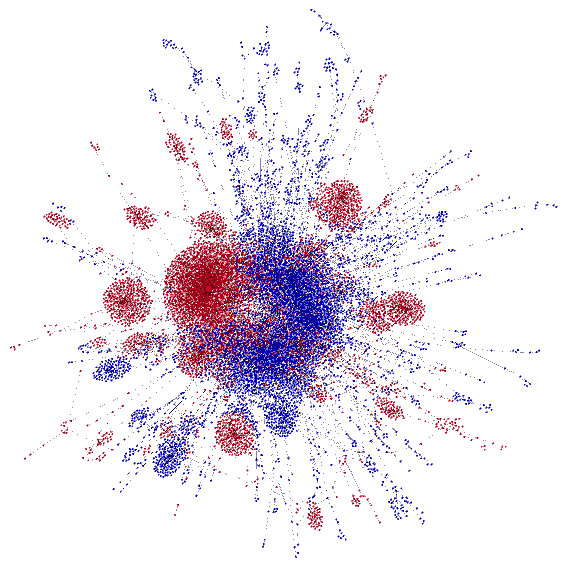}
\caption{Force directed layout visualization of a non-controversial topic (\#sxsw).}
\end{center}
\label{fig:sxsw}
\end{marginfigure}

As explained in the previous section, we use METIS~\cite{karypis1995metis} to produce two partitions on the {retweet graph}.
Given a retweet graph and its two partitions, we produce a visualization of the graph, as in the cases of Figures~\ref{fig:russia_march} and \ref{fig:sxsw}.
The figures display the two partitions for two topics (\#russia\_march and \#sxsw) in blue and red color on their corresponding retweet graphs.
The graph layout is produced by Gephi's ForceAtlas2 algorithm~\cite{jacomy2014forceatlas2}, and is based solely on the structure of the graph, not on the partitioning by METIS.
It is easy to see that the retweet graph of the first topic is characterized by a bi-modal clustering structure, indicating a controversy.
In contrast, the retweet graph of the second topic is characterized by a uni-modal clustering structure, indicating lack of controversy.

%\input{demo_description.tex}  
% !TEX root = controversy.tex
\section{Exploration Session}
\label{sec:demo}

The demonstration will allow the attendees to explore the set of trending topics discussed on Twitter during Jun-Sep 2015.
The attendees will be able to interact with a web interface to select one topic and retrieve its \emph{summary}, and organize the set of topics according to different \emph{views}.

\spara{Topic Summary.}
The summary of each topic consists of its hashtag, together with the most related keywords, which convey the main idea behind the topic itself.
To further help in understanding the topic, the system provides representative tweets from either side of the % (possible) 
controversy.
These tweets come from authoritative vertices in the graph, as measured by the number of endorsement received.
Finally, the summary also includes a visualization of the retweet graph.

\begin{figure}[t]
\centering
\includegraphics[width=\columnwidth]{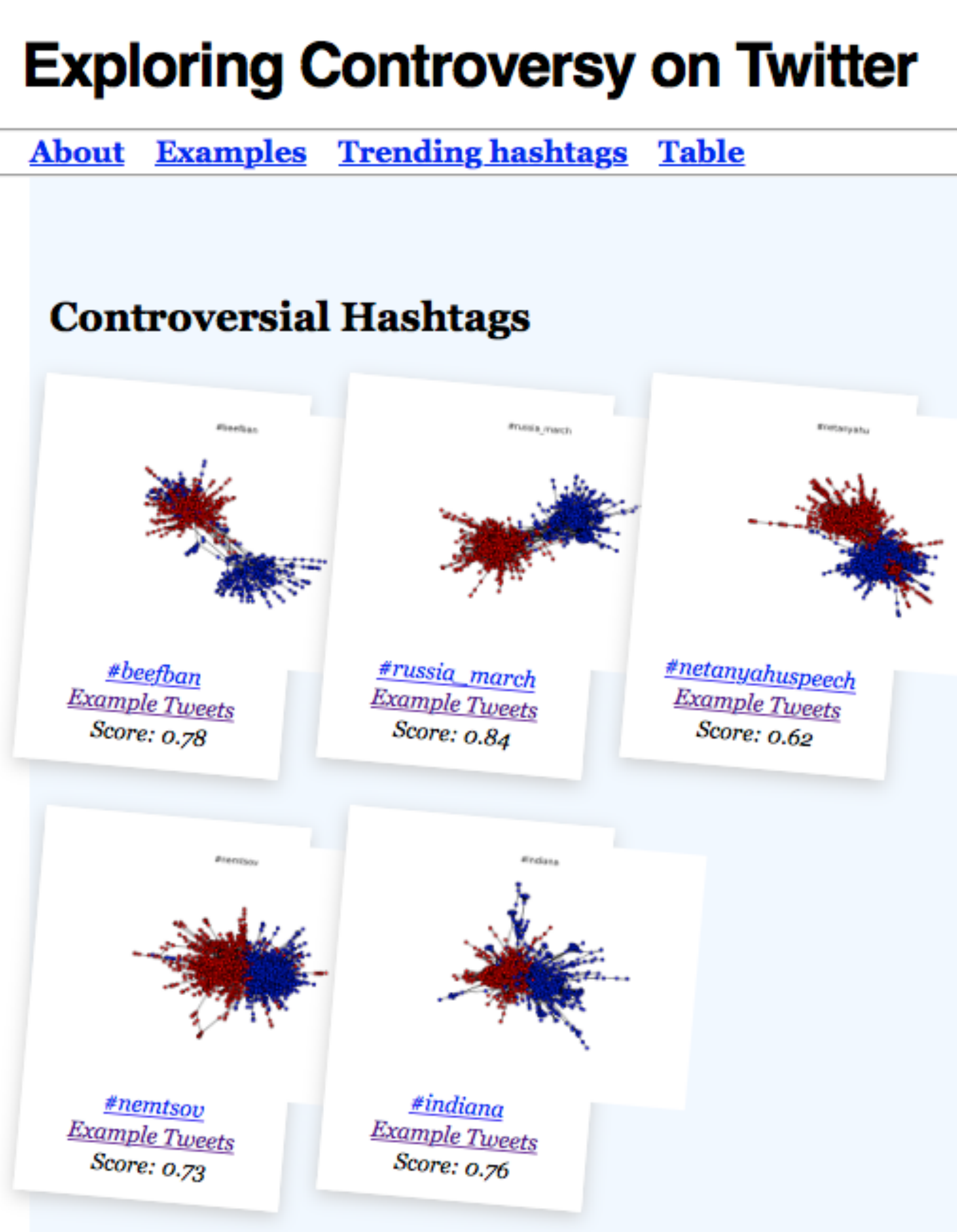}
\caption{Screenshot of the web interface.}
\label{fig:screenshot}
\end{figure}

\spara{Topic Views.}
Attendees will have the option to browse the topics in chronological order, or sorting them by controversy score, to find the most controversial ones.
The system offers also a search functionality, by which the user can specify a text query and obtain a set of relevant topics.
Figure~\ref{fig:screenshot} is a screenshot of the system while showing some examples of the most controversial topics.

\balance{}
\bibliographystyle{SIGCHI-Reference-Format}
\bibliography{biblio}

\end{document}